\documentclass{ws-p8-50x6-00}

\def\stacksymbols #1#2#3#4{\def\theguybelow{#2}
\def\verticalposition{\lower#3pt}
\def\spacingwithinsymbol{\baselineskip0pt\lineskip#4pt}
\mathrel{\mathpalette\intermediary#1}}
\def\intermediary#1#2{\verticalposition\vbox{\spacingwithinsymbol
\everycr={}\tabskip0pt
\halign{$\mathsurround0pt#1\hfil##\hfil$\crcr#2\crcr
\theguybelow\crcr}}}

\def\gapprox{\stacksymbols{>}{\sim}{4}{1}}
\def\lapprox{\stacksymbols{<}{\sim}{4}{1}}

\def\simlt{\lower.5ex\hbox{$\; \buildrel < \over \sim \;$}}
\def\simgt{\lower.5ex\hbox{$\; \buildrel > \over \sim \;$}}
\def\simpropto{\lower.2ex\hbox{$\; \buildrel \propto \over \sim \;$}}

\begin{document}

\title{{THE DARK SIDE OF THE UNIVERSE}
\footnote{To appear in  {\em 2001: A Spacetime Odyssey} (World Scientific,
Singapore)}
}
\author{Joseph Silk}

\address{Department of Astrophysics,  Keble Road,\\
Oxford OX1 3RH, England\\E-mail: silk@astro.ox.uk}

\maketitle

\abstracts{Most of the matter in the universe is invisible.
I review the status of  dark matter and describe how
both the theory of galaxy formation and novel types of experimental
searches are   revitalizing
attempts to find non-baryonic dark matter.
}

\section{Introduction}

This is an interesting time for dark matter studies. Direct and indirect
searches will soon be capable of exploring most of the allowable parameter
space for the elusive stable SUSY relics that are the favoured CDM
candidate. Gravitational lensing is mapping the distribution of dark matter
in halos and over larger scales with unprecedented precision.

The nature of the dark matter still
almost completely eludes us. Nevertheless, there are
many conjectures and proposed searches for dark matter signatures. In this
review, I  present an historical summary of the dark matter
problem, review our present understanding of the distribution of dark matter
and the various candidates for baryonic and non-baryonic dark matter, and
describe the implications of particle dark matter for galaxy formation.  
\section{A brief history of dark matter probes}
\subsection {Galaxy clusters}
The quantitative case for dark matter was first made by Zwicky\cite{zwi} in
1933 for the Coma cluster of galaxies. He used galaxy peculiar
velocities to estimate the virial mass, which he compared with the stellar
mass, estimated from the luminosity. This method is essentially unchanged
today, and modern applications of the viral theorem yield a similar value
of mass to luminosity, of about 300 $\rm M_\odot/L_\odot.$ However the
uncertainties are large. Optical measurements utilizing the x-ray luminosity
yield the emission measure and temperature from which one may derive the
pressure gradient, and at submillimeter wavelengths, the Sunyaev-Zeldovich
effect directly yields the electron density. These give similar values of
$M/L,$ but again, the uncertainties are at least 50 percent.

Gravitational
lensing has provided a direct measure of dark mass, determining the total
surface density within the Einstein radius.  These measurements make no
assumptions about virialization of the gas, any deviations from
isothermality, sources of nonthermal gas or velocity anisotropy. 
There are 
 already hints 
 from some large-scale shear studies that the inferred mass-to-light ratios,
 may differ from those inferred from large redshift surveys.
 For example, it is claimed that (Wilson, Kaiser and Luppino
2001)\cite {wil}
the early-type
 galaxies trace the mass density at $z\sim 0.5$, resulting in  $\Omega_m\approx 0.1$
if any contribution from late-type galaxies is neglected.  Weak lensing studies of groups (Hoekstra et al. 2001)\cite{hoe} 
 also lead to a relatively low $(M/L)_B=183\pm 80 h$ at $z=0$,
about half the value typically found for rich 
 clusters (e.g. Carlberg et al. 1997).\cite{car}
Independent techniques for inferring cluster masses exist, but
usually involve additional assumptions. For example, deduction of masses from
studies of x-ray emission from the diffuse intracluster gas usually
requires such assumptions as hydrostatic equilibrium of the
gas and neglect of non-thermal support.
This result is  quantified in a study of x-ray clusters, which finds that  
$ (M/L)_v\approx 170 (T/1\rm keV)^{0.3}$ (Bahcall and Comerford 2001).\cite{bahc}

There is an interesting corollary of this result.
If $\Omega_m\approx 0.3,$ as many
large-scale studies, { e.g.} of redshift-space distortions, suggest, then
$(M/L)_v \approx 300$ for typical regions of the universe that are
 unbiased, adopting $(M/L)_v \approx 1400\Omega_mh$ as the canonical normalisation.
This supports the usual assumption that the scale of galaxy clusters
probes  such representative regions,
despite the fact that clusters predominantly contain early-type galaxies.

The reason for the differences  between the different environments
might be due to a systematic  decrease of $M/L$ in less dense clusters,
in the outer regions of rich clusters and in the field,
relative to the cluster
 value. The differences are   
indeed likely to be associated with  galaxy type: the cores of rich clusters
are dominated by spheroid-dominated (early-type) galaxies,
but the outer parts of clusters and the field are dominated 
by disk-dominated (late-type) galaxies.
Nevertheless, the trend in cluster
mass-to-light ratios is most simply understood if the efficiency of galaxy
formation at incorporating baryons into stars in galaxies 
in  low density environments
is about
twice that for galaxies in  massive clusters. 

There may  be an increase  of
cluster $M/L$ with increasing cluster  central density
because  denser clusters are older and 
would have undergone more passive evolution. However any
significant evolution, e.g. as much as a magnitude,  might be hard to
 reconcile with the similar fundamental plane relation  (and hence $M/L$
range) 
found for early-type field and cluster galaxies  (van Dokkum et al. 2001).\cite{dok}

Since approximately two-thirds of  the 
optical ($B$-band) light density of the universe is produced
by disk-dominated (late-type) galaxies,
it follows, taking these $M/L$ estimates at face value,
that these galaxies are  associated,
presumably via their dark halos, with about a third  of the mass in the universe.
It is therefore not surprising that early-type galaxies trace the mass.

\subsection{Galaxy halos}

In the 1960's, the dark matter content of spiral galaxies was extensively
explored by Rubin at optical wavelengths, using H$\alpha$ radial velocity
measurements of HII regions and by Roberts at 21cm, using the diffuse
atomic interstellar medium as a dynamical probe of radial velocities. 
Modern observers have used rotation curves to explore the decomposition of
the inner parts of galaxies into disk, bulge and dark halo contributions.
Within the optical region (e.g. the half-light radius or typically several kpc),
up to half of the matter contributing to the rotation curve may be
non-baryonic (but see discussion below).  Disks extend to 50 kpc or more for typical spirals and
non-baryonic matter accounts for $\sim 90$ percent of the dynamical mass.
  The
inferred total $M/L$ ratio is typically in the range 10 to 50.  
A self-similar profile with
only two free parameters, central density and concentration,  fits most dark
halos.

  Low surface brightness spirals
provide  an exception to the universal profile
fit. These 
systems
are everywhere dark-matter dominated and display a variety of central cores
 which are often soft and rarely cuspy
 (van den Bosch and Swaters 2001).\cite{van}

\subsection{Galaxy redshift surveys}

On  very large scales, from 10 to 100 
Mpc, progress has awaited the results
from the large redshift surveys.  The pioneering CfA surveys of the 1980s
and 1990s with some 8000 redshifts culminated
in the Las Campanas survey of 25000
galaxies in the mid 1990s.  The situation changed little  until the AAT 2DF ushered in the first
250,000 galaxy survey in
2001, followed closely by SDSS with the prospect of $10^6$
redshifts by 2003.
The universe is well-sampled to $z\sim 0.3.$

An important result is that redshift space distortions allow measurement of
the dark matter content, subject to the unknown bias of mass relative to
light. If the bias is assumed to be scale-independent and described by
 a spectrum of density fluctuations with variance
$$\sigma^2=\langle(\delta\rho/\rho)^2\rangle/\langle(\delta\rho/\rho)^2\rangle_{galaxies},$$
normalized to a fiducial scale of $8h^{-1}\rm Mpc$ where
$\langle(\delta\rho/\rho)^2\rangle_{galaxies}\approx 1,$ one infers via
 deviations from
the Hubble flow as measured by redshift-space distortions that (Peacock et
al. 2001)\cite{pea}
$\Omega^{0.6}\sigma_8= 0.43\pm 0.07.$ 
An independent approach uses the shape of the power spectrum,
combining the 2DF galaxy redshift survey with
cosmic microwave background fluctuations, to obtain a virtually identical result
(Efstathiou et al. 2001).\cite{efs}
The inferred global $M/L$ is
$300\sigma_8^{-1.7}$.  Theory suggests that on scales above 10 Mpc which are
sampled by the large-scale surveys, one would anticipate little or no bias
$(\sigma_8\approx 1)$. Hence $\Omega_m\approx 0.3.$

\subsection{The baryon fraction}

Independent measures of $\Omega_m$ confirm that $\Omega_m \approx 0.3$, for
example both via the evolution of the cluster number density
 (Bahcall and Fan)\cite{bahf}
and  the observed cluster baryon fraction (Arnaud and Evrard
1999)\cite{arn}. 
The universal primordial baryon fraction is well measured from light element
nucleosynthesis to be 0.05 (adopting a Hubble content of 65 km/s/Mpc).  The
baryon fraction of 15\% which is actually measured in rich galaxy clusters
 requires the universal dark matter density to be $\Omega_m\approx
0.3,$ 
and confirms the result inferred from large-scale structure studies.

However  a   baryon fraction of 15\% can only be reconciled with galaxy
halo and disk masses, and galaxy formation, if while beginning with
$\Omega_b/\Omega_m\approx 0.15, $  one ends up with $\Omega_\ast/\Omega_m\approx 0.07,$ where $\Omega_\ast$ is the observed  luminous baryon density at $z=0.$
This constraint comes from several independent arguments, and  one requires 
about 50 percent of the
baryons to either be ejected or to hitherto be undetected. 

The Milky Way mass budget suggests that 
 the galactic
disk and spheroid amount to 
$7\times 10^{10}\rm M_\odot$ in a halo of about  $10^{12}\rm M_\odot.$ 
A more quantitative confirmation of this
argument comes from examining the adiabatic compression 
of the dark halos by cooling baryons, both analytically
in comparison with the velocity function
(Kochanek and White 2001; Kochanek 2001)\cite{kocw,koc}
 and via  numerical 
 simulation modelling of the Milky Way in comparison with the rotation curve
(Klypin, Somerville and Zhao 2001).\cite{klyp} The inference that more baryons
cooled than are seen in stars is a 
manifestation
of the so-called cooling catastrophe.
There appears to be  a similar shortfall in the local mass in old stars as seen in $K$-light compared to the rest-frame  ultraviolet luminosity density
observed in star-forming galaxies at $z\sim 3$,
for reasonable extinction corrections (Cole et al. 2001).\cite{cole}

The obvious  solution to this problem involves either 
expelling half the  the gas or hiding the gas  before it forms stars.
Both hypotheses are 
 controversial since supernovae seem incapable of driving
so much gas out of the early galaxy (MacLow and Ferrara 1999)\cite{mac}
and there is little evidence for a
substantial halo component of dark baryons
either in diffuse gas, which would have been detected,
or in compact form, as in MACHOs, over the mass range $10^{-7}\lapprox M_{MACHO}\lapprox
 10\rm M_\odot$ (Milsztajn and Lasserre 2000).\cite{mill}

\section{What is the dark matter?}
 There are two distinct varieties of dark matter.
There is both  baryonic dark matter and non-baryonic matter.
Each provides distinct difficulties.

\subsection{Baryonic dark matter}
 Primordial
nucleosynthesis has convincingly predicted the baryon fraction of the
universe to be $\Omega_b \approx 0.05$. There is an independent estimate of
$\Omega_b$ from the fluctuations in the cosmic microwave background, and in
particular from the height of the second acoustic peak. One needs to assume a
cosmological model,
such as that of a flat universe, and the predominance of the
adiabatic mode of primordial fluctuations for this measurement to be entirely free of
degeneracy. However  flatness is unambigously measured from the location of the first peak, and inflationary cosmology guarantees 
the emergence of adiabatic fluctuations from quantum fluctuations in the presence of a single scalar inflaton field.

Direct observations
of the baryon fraction that utilize the luminous components of galaxies and
the absorption of the diffuse intergalactic medium give values of
$\Omega_b$ that at $z\approx 3$ agree with the primordial nucleosynthesis
value (most notably via Lyman alpha forest absorption) but at $z\approx 0$ fall short by
a factor of around 4.
 The only reasonable conclusion is that there are dark
baryons:  75 percent of the baryonic component has somehow avoided the
process of luminous galaxy formation or of being incorporated
into
the cold/warm component of the intergalactic gas that is associated
with the Lyman alpha absorbing clouds detected at low $z$.

A diffuse, warm
intergalactic medium component at $10^5$ to $10^6$K may account for some of
the baryonic dark matter.  However, the numerical simulations which predict
its existence and heating via gravitational accretion shocks around galaxies
and galaxy clusters make a plausible case for
at most  only 30-40 \% of the
baryonic dark matter to be in this form.

One might expect that  local processes of
star formation selected the baryons seen in stars, leaving the remaining
baryons in the vicinity of the stellar components. This is not necessarily
the case, however, as the Lyman alpha forest tracers of high
redshift baryons are only weakly correlated, or even anti-correlated, with
respect to luminous galaxies.
Indeed, the nearby $(z\simlt 0.1$) Lyman alpha absorbing clouds account for some 20-25\% of the baryons in low column density warm, photo-ionized gas (Shull 2001).\cite{shu}
With say 10\% in stars,
 at least  25\% of the baryons are unaccounted for.

Presumably these
``missing''
dark baryons are
today in galaxy halos. 
In the absence of any further indications of cold intergalactic baryons, searches have therefore  focussed  on dark
halo baryons.
Halo searches have failed to come up
with dominant amounts of baryonic dark matter.  
The best motivated candidate,
MACHOs,  amount to no more than
20 percent of the halo mass between the solar circle and the LMC
(Alcock et al. 2000).\cite{alc} 
However this is almost enough to account for the remaining cooled baryons:
if the MACHO limit is saturated, there is about twice as much mass in MACHOs as in stars.

Of course 
MACHOS remain the only claimed detection
although their halo distribution is controversial.  If indeed the MACHOs are
a halo component, then old white dwarfs provide a possible MACHO candidate.
While the MACHO experiment does permit a significant  
halo mass fraction in the form of stellar remnants,  overproduction of chemical elements 
strongly argues against the interpretation of all of the MACHOs as being
white dwarfs (Fields, Freese and Graff 2001).\cite {fie}
The claimed detection of halo white dwarfs at the few percent level
(Oppenheimer et al. 2000)\cite{opp} has been strongly criticized on
kinematic grounds, 
but chemical abundance anomalies in old halo stars and in extragalactic
deuterium are  consistent  with the proposed  abundance  of halo
white dwarfs and with the existence of
 the inferred primordial population of intermediate mass
stars (Fields et al. 2001).\cite{fieo}

 Alternative halo baryonic dark matter candidates are elusive.
Cold $H_2$ clumps are an intriguing possibility since such gas would have
evaded detection in a halo component.  There may be a large H$_2$ reservoir:
indeed it has been argued that the HI in the interstellar medium is
photodissociated H$_2$ (Allen 2001).\cite{all}
 However the observational hints of extensive H$_2$
that support this hypothesis suggest that the total mass detected in this
form in the inner galaxy is unlikely to exceed the cold HI mass.  Another
(admittedly theoretical) difficulty with this argument
 is that while there is some evidence for
diffuse cold interstellar H$_2$, cold H$_2$ clumps pose a stability problem:
why don't they form stars?

\subsection{Non-baryonic dark matter}
Baryonic dark matter only accounts for between 10 and 25 percent of the dark
matter.  The nature of the non-baryonic dark matter is unknown. However SUSY
has provided motivated candidates among the class of the lightest stable
particles, generically called neutralinos. This stems from the remarkable
coincidence that with $\Omega_m \approx 0.3$, as required by large-scale
structure observations, the particle annihilation cross-section is required
via thermal freeze-out at $T\approx m_x/20$ to satisfy
$\langle\sigma v\rangle_{ann}\approx 3 \times 10^{-26}(0.1/\Omega_m h^2)\rm cm^3s^{-1},$ within the range
expected for typical WIMP candidates.  The prime uncertainty in translating
this constraint into WIMP search parameter space is that even minimal SUSY
models allow a range of several orders of magnitude in the cross-section
predicted for a specified WIMP mass $m_x.$ Accelerator constraints impose the
lower bound
 $m_x\gapprox 50-100\rm GeV$ (Baltz and Edsjo 2001;
Ellis, Nanopoulos and Olive 2001).\cite{balg,eno}
Hence annihilation signatures are potentially
detectable 
by high energy astrophysics experments that search for diffuse halo
gamma rays, neutrinos or even cosmic ray fluxes generated in the halo.

High resolution simulations provide a new approach to studying
weakly interacting dark matter. Halos are found to be clumpy,
the substructure persisting through successive mergers and not necessarily
being fully resolved by the best simulations to date.
One consequence of the substructure is a large population of satellite
galaxies,
approximately increasing in abundance  as $dN/dM\propto M^{-2}.$
In fact the low observed 
frequency of dwarf satellites  argues
against a halo that is as highly clumped as suggested by the CDM simulations.
Independent confirmation of this local result comes from
high redshift observations  of the gas-rich pregalactic phase  provided
by the data on damped Lyman alpha clouds seen towards quasars.
A recent survey (Prochaska and Wolfe 2001)\cite{pro} fails to find  the predicted excess of low line width  clouds relative to clouds with large $(\sim 200\rm km \,s^{-1})$ velocity dispersions.

Quenching of star formation by early $(z\sim 10)$ photoionization may help make the
dwarfs invisible (Gnedin 2000; Somerville 2001)\cite{gne,som}.
Overheating of the disk could still pose a problem, although 
accounting for the observed 
 thinness of the galactic disk may be  less of a difficulty 
than previously suggested if $\Omega_m$ is low
(Font and Navarro 2001)\cite{fon}.

The simulations also predict excessive concentration of the dark matter,
cuspiness of the halos, and loss of angular momentum via dynamical friction,
all of which seem to result in signatures that appear to contradict the
observational evidence from studies of the stellar component.

Several indirect observational probes suggest that in addition to the substructure issue, the dark matter
concentration and cuspiness  are rather less extreme than
predicted by the simulations. 
The high resolution simulations generally find a dark matter profile with a central cusp $\rho \simpropto r^{-1.5}$ for galaxy mass halos (Moore et
al. 1999; Jing and Suto 2000;
Klypin et al. 2001).\cite{moo,jing,klypa}
However observations of dwarf LSB spirals find little,
if any, evidence for central cusps.

The predicted concentration
for the typical dark halo in a $\Lambda CDM$ model results in about a 50\%
dark matter contribution within 2 disk scale lengths. 
 With regard to Milky-Way type
galaxies, the evidence suggests that dark matter cusps do not exist in the presence of
bars, both from dynamical arguments centering on dynamical friction of the rapidly rotating bars on the dark matter
that require the baryons in bars to
be self-gravitating (e.g., Debattista and Sellwood 2000)\cite{wei}
and the associated disk to be maximal, and from the microlensing and stellar population
modelling of the inner Milky Way in combination with the observed rotation
curve (Binney and Evans 2001).\cite{bin} This latter constraint limits the
axially symmetric non-baryonic dark matter contribution within the solar
circle to be less than or of order 10 percent.  Of course the Milky Way has a
bar: however one well-studied
nearby spiral with no evident bar appears to require  a  $\sim 50$\%  dark
matter contribution within the optical disk for a submaximal disk model
(Kranz, Slyz and Rix 2001) to account for features in the measured rotation curve.\cite{kra}

Tilting the primordial spectrum to the red
lowers the small-scale power, and lowering of $\Omega_m$ reduces the concentration of halo dark matter.
However  these patchwork solutions
do not  completely alleviate the substructure and concentration problems,
and 
a red tilt is likely to create other difficulties.

We do not know if the resolution of these issues lies in the domain of
fundamental gravity, particle physics via 
tinkering with the nature of the dark matter particles  or  astrophysics via modification of the dark halo
properties.
Fundamental changes in the gravity law
may allow one to modify Newton's laws, either by a phenomenological  approach
that  fits rotation curves with one additional parameter and no dark matter, but is not Lorentz invariant (Milgrom 1999)\cite{milg},
or by invoking  higher-dimensional gravity 
to introduce gravitational interactions with  adjacent branes
(e.g., Arkani-Hamed et al. 2000)\cite{arkh}

 Adjusting the neutralino properties, { e.g.} by allowing 
the particles  to be
self-interacting or fluid-like, modifies the halo properties, although not necessarily in
a completely satisfactory direction (Yoshida et al. 2000a;
Yoshida et al. 200b;
Meneghetti et al. 2001).\cite{yosa,yosb,men} 
Warm dark matter does not solve the cusp problem
(Avila-Reese et al. 2001; Knebe et al. 2001)\cite{avi,kne}, 
although the substructure problem is largely
resolved (Bode, Turok and Ostriker 2001).\cite{bod}
The median satellite distances from  the Milky Way,
already in disagreement with $\Lambda CDM$ model predictions,
are considerably aggravated by warm dark matter (Klypin 2001,
private communication).
More complex modifications of particle matter have  been
invoked. 

Alternatively, early dynamical processes involving non-axially
symmetric distributions of baryons could couple the baryons to the dark
matter, erode some of the cuspiness and substructure,
and reduce the concentration. This may happen dynamically via black hole mergers
or via a combination of the  deceleration of a rapidly rotating bar
and black-hole driven outflows
 (Binney, Gerhard and Silk 2001).\cite {bings} This 
latter approach is more difficult to
simulate and quantify, but some
dynamical modification of the dark matter distribution 
clearly must occur in the
presence of bar formation and dissolution, supermassive black hole formation,
and massive outflows.

One can imagine the following sequence of events.
A merger between protogalaxies results in formation of a  massive, rapidly rotating, transient bar. The non-axially symmetric matter distribution exerts strong tidal torques on the gas that is driven into the centre of the galaxy to form a supermassive black hole.
Meanwhile the dynamical friction exerted by the bar heats 
and spins up
the halo core within a region containing roughly equal amounts
of baryons and dark matter. The bar contracts before it dissolves
as the black hole grows, and the halo core expands. The inner galaxy is 
now baryon-dominated.
Outflows driven by accretion of gas onto the supermassive black hole sweep out
substantial amounts of baryons, of order  50\%, until the 
core contains roughly equal amounts of dark matter and baryons,
when baryon-driven growth of the black hole becomes less important.
The outflows consist of low angular momentum baryonic matter.
The disk forms by late  infall of high angular momentum baryons.

This suggests that all spheroids underwent an early quasar-like phase.
The following problems {\it may} be solved: the dark matter cusp is erased, the concentration is reduced,  dwarfs are stripped of gaseous baryons, baryons are ejected in a substantial amount, and disk sizes are enhanced because disks form 
from high angular momentum gas.
While these claims are meant only to be taken qualitatively, the point is clear: astrophysical processes may strongly modify the dark halo properties.

\section{``Observing'' non-baryonic dark matter} 

Dark matter particles are majorana-like, and their annihilations provide a
potentially
powerful signature.
There have been tentative reports of high energy $e^+$ and diffuse high
galactic latitude gamma ray detections.  However the fluxes measured are up
to 100 times in excess of what would be predicted for a uniform CDM halo
(Baltz and Edsjo 1999; Baltz et al. 2001)).\cite{bal,bala} 
In
fact, halos are plausibly clumpy, as revealed by high resolution simulations
of halo formation, but it is not known whether the clumpiness persists until
the present epoch and suffices to account for the enhanced fluxes.  

Ideally, one would like a direct probe of the dark matter.  This may be
provided within our galaxy or in nearby galaxies by searching for the annihilation signal which will be enhanced by
the concentration of CDM towards the inner halo, and in particular by the
possibility of a central cusp.  Again, such cusps are predicted by the
simulations to have $\rho\propto r^{-1.5}.$ In fact,
 not only are such cusps not
seen, but they would not significantly enhance the annihilation signal.

However the situation changes dramatically in the presence of a central
supermassive black hole (SMBH). Such objects are 
virtually ubiquitous in spheroids, and
their mass correlates tightly with spheroid velocity dispersion over the 
entire SMBH
mass range, with the median
black hole mass fraction equal to 0.13\% that of the bulge
(Ferrarese and  Merritt 2001;
Gebhardt et al. 2001).\cite {fer,geb}
The supermassive black holes  range in mass
from the SMBH in
the Milky Way bulge  ($\sim 
2.6\times 10^6\rm M_\odot$) to NGC 4258 ($\sim 
4\times 10^7\rm M_\odot$), these  being the two
unchallengeable supermassive black hole cases,
  up to M87 ($\sim 
3\times 10^9\rm M_\odot$). 

 There are two effects. Firstly, the halo forms via mergers of
subsystems containing smaller SMBHs. This is expected in hierarchical halo
formation.  As the smaller SMBH spirals into the dominant halo, the inner
halo cusp is smoothed, to a profile that, for minor mergers, approaches $\rho\simpropto r^{-0.5}$
(Nakano and  Makino 1999; Merritt and Cruz 2001)\cite{nak,mer}.
For major (equal mass) mergers,
a steeper profile results,  $\rho\simpropto r^{-1}$ (Milosavljevic
and Merritt 
2001).\cite{mil}

Stellar cusps have been studied in ellipticals.  
Power-law cusps are found in faint spheroids and in rotating  systems, while luminous
ellipticals often have central soft cores
that are inferred to have been generated dynamically 
by  the orbital decay of 
the central SMBH (Faber et al. 1997).\cite{faber} 
A range of  dynamical histories is inferred that must depend on whether the SMBH grew as a result of a dynamical merger or by gas accretion.

CDM continues to accrete while the central SMBH grows by accreting more mass
via baryonic dissipation in the early quasar outburst, associated with,
and driven by, gas
infall onto the SMBH, that occurs during the gas-rich protogalactic phase.
The dominant process underpinning the bulk of the SMBH growth
 must be baryonic dissipation,
as dynamical growth by black hole or stellar mergers is simply too slow to
produce a significant space density of ultraluminous quasars at $z\simlt 6,$ as
observed.  This is especially true for disk galaxies
where significant black hole growth by  black hole mergers
following mergers of galaxies containing black holes would 
probably have been overly disruptive of the fragile disks.

 The Milky Way galaxy, with a relatively insignificant spheroid, 
is a candidate for SMBH growth by accretion and a dense dark halo
cusp. Any mergers in the Milky Way must have occurred long ago, to avoid disk disruption, and the SMBH is likely to have acquired its present mass of $2.6\times 10^6 \rm M_\odot$ as a result of gas accretion.
There is even now a large reservoir of gas in the galactic centre, both in molecular gas and in the inferred ejecta from OB stars, although the current accretion rate of the SMBH is extraordinarily low at the present epoch.
The result is that, almost certainly in our own galaxy,  the CDM density 
 around the SMBH is locally
enhanced by the adiabatic response of the CDM to the SMBH growth. 

 Whatever the original dark matter core, the central cusp  becomes denser
and steeper as a consequence of quasi-adiabatic SMBH growth.
The inner
cusp develops within the zone of gravitational influence of the SMBH,
$GM_b/\sigma^2,$ of order a parsec for the Milky Way.  This inner cusp,
$\rho\propto r^{\gamma\prime},$ is always steeper than $\rho\propto
r^{-1.5},$ for $\it any$ underlying halo profile $\rho\propto r^\gamma:$
$\gamma\prime=\frac{9-2\gamma}{4-\gamma},$
and yields observable potentially fluxes of high energy neutrinos (Gondolo
and Silk 1999).\cite{gon}

The resulting  signal would be divergent,
 except that sufficiently close to the SMBH
(within $< 100$ Schwarschild radii), the annihilation time becomes so short
that the CDM density declines abruptly.
Thus  a central supermassive black hole  further enhances the
annihilation signal by concentrating dark matter during the halo
formation phase. The predicted signals 
(Bertone, Sigl and Silk 2001, and in preparation)\cite{ber} include coincident point-like radio and  gamma
ray emission which are consistent with observations of 
SagA* with regard to both flux and spectral signatures. 
Future observations of terrestrial muon fluxes induced by
high energy neutrinos propagating through the earth, for example  with 
deep underwater detectors such as ANTARES, should help
test this conjecture.

\section*{Acknowledgments}I thank Gianfranco Bertone, Anatoly Klypin 
and Guenter Sigl for
 fruitful discussions and ongoing collaborations.

\end{document}